\documentclass[10pt,journal,epsfig]{IEEEtran}

\usepackage[dvips]{graphicx}
\usepackage{graphicx}
\usepackage{amssymb}
\usepackage{cite}
\usepackage{subfigure}
\usepackage{amsmath}

\begin{document}

\title{Prior Support Knowledge-Aided Sparse Bayesian Learning with Partly Erroneous Support Information}

\author{Jun Fang, Yanning Shen, Fuwei Li, and Hongbin Li,~\IEEEmembership{Senior
Member,~IEEE}
\thanks{Jun Fang, Yanning Shen and Fuwei Li are with the National Key Laboratory
of Science and Technology on Communications, University of
Electronic Science and Technology of China, Chengdu 611731, China,
Email: JunFang@uestc.edu.cn}
\thanks{Hongbin Li is
with the Department of Electrical and Computer Engineering,
Stevens Institute of Technology, Hoboken, NJ 07030, USA, E-mail:
Hongbin.Li@stevens.edu}
\thanks{This work was supported in part by the National Science
Foundation of China under Grant 61172114, and the National Science
Foundation under Grant ECCS-1408182. }}

\maketitle

%theoretically and experimentally

%

%is able to seamlessly integrate the prior support information into
%the conventional sparse Bayesian learning framework

%we first introduce a slightly modified hierarchical model which,
%through placing a different value to a model-related parameter,
%can assign non-sparsity-encouraging priors to those coefficients
%which are believed to be nonzero.

%The proposed modeling constitutes a three-layer hierarchical form.

%The first two layers, similar to the conventional sparse Bayesian
%learning, place a Gaussian-inverse-Gamma prior on the signal,
%while the third layer is newly added to the conventional sparse
%Bayesian learning framework.

%Instead of using a common parameter for all sparsity-controlling
%hyperparameters $\{\alpha_i\}$,

\begin{abstract}
It has been shown both experimentally and theoretically that
sparse signal recovery can be significantly improved given that
part of the signal's support is known \emph{a priori}. In
practice, however, such prior knowledge is usually inaccurate and
contains errors. Using such knowledge may result in severe
performance degradation or even recovery failure. In this paper,
we study the problem of sparse signal recovery when partial but
partly erroneous prior knowledge of the signal's support is
available. Based on the conventional sparse Bayesian learning
framework, we propose a modified two-layer Gaussian-inverse Gamma
hierarchical prior model and, moreover, an improved three-layer
hierarchical prior model. The modified two-layer model employs an
individual parameter $b_i$ for each sparsity-controlling
hyperparameter $\alpha_i$, and has the ability to place
non-sparsity-encouraging priors to those coefficients that are
believed in the support set. The three-layer hierarchical model is
built on the modified two-layer prior model, with a prior placed
on the parameters $\{b_i\}$ in the third layer. Such a model
enables to automatically learn the true support from partly
erroneous information through learning the values of the
parameters $\{b_i\}$. Variational Bayesian algorithms are
developed based on the proposed hierarchical prior models.
Numerical results are provided to illustrate the performance of
the proposed algorithms.
\end{abstract}

%such that allows to automatically learn the true support from the
%partly erroneous knowledge.

%to learn the parameters $\{b_i\}$ along with the sparse signal

%where $\{b_i\}$, along with $a$, are parameters characterizing the
%Gamma hyperpriors imposed on the sparsity-controlling
%hyperparameters $\{\alpha_i\}$.

\begin{keywords}
Compressed sensing, sparse Bayesian learning, prior support
knowledge.
\end{keywords}

%A new sparse Bayesian learning method, also referred to as partial
%support knowledge-aided sparse Bayesian learning, is developed to
%automatically learn the correct support from the partly erroneous
%knowledge and to recover the sparse signal.

%i.e. the presumed partial support includes a set of correct
%estimates and a set of incorrect estimates of the nonzero
%coefficients' locations of the sparse signal

%In other words, the presumed partial support includes a correct
%part which is the true support of the sparse signal, and an
%incorrect part which is not the support of the sparse signal but
%treated as so.

%In practice, the knowledge of the support of the sparse signal may
%be obtained from the estimate of the previous time instant.

%Note that $\{b_i\}$, along with $\{a_i\}$, are used to
%characterize the Gamma priors placed on the hyperparameters
%$\{\alpha_i\}$. These parameters are set to be very small values
%in the conventional sparse Bayesian learning modeling.

%A nonzero parameter $b_i$ encourages a small $\alpha_i$ and thus
%does not have the potential to promote the sparsity of the
%corresponding coefficient $x_i$.

\section{Introduction}
Compressed sensing is a recently emerged technique for signal
sampling and data acquisition which enables to recover sparse
signals from much fewer linear measurements
\cite{ChenDonoho98,CandesTao05,Donoho06,TroppGilbert07,Wainwright09}.
In this paper, we study the problem of sparse signal recovery when
prior information on the signal's partial support is available. In
practice, prior information about the support region of the sparse
signal may come from the support estimate obtained during a
previous time instant. This is particularly the case for
time-varying sparse signals whose support changes slowly over
time. For example, in the real-time dynamic magnetic resonance
imaging (MRI) reconstruction, it was shown that the support of a
medical image sequence undergoes small variations with support
changes (number of additions and removals) less than $2\%$ of the
support size \cite{VaswaniLu10}. Also, in the source localization
problem, the locations of sources may vary slowly over time. Thus
the previous estimate of locations can be used as prior knowledge
to enhance the accuracy of the current estimate.

The problem of sparse signal recovery with partial support
information was studied in several independent and parallel works
\cite{Vaswani10,VaswaniLu10,KhajehnejadXu09,MiossoBorries09}. In
\cite{Vaswani10}, the partly known support is utilized to obtain a
least-squares residual and compressed sensing is then performed on
the least-squares residual instead of the original observation.
When the partly known support is accurate, it is expected that the
sparse signal associated with the residual has much fewer large
nonzero components. Hence the least-squares residual-based
compressed sensing can improve the recovery performance. Later in
\cite{VaswaniLu10}, a weighted $\ell_1$-minimization (modified
basis pursuit) method was proposed, where the partially known
support is excluded from the $\ell_1$-minimization (equivalent to
setting the corresponding weights to zero). Sufficient conditions
for exact reconstruction were derived, and it was shown that when
a fairly accurate estimate of the true support is available, the
exact reconstruction conditions are much weaker than the
sufficient conditions for compressed sensing without utilizing the
prior information \cite{VaswaniLu10}. This work was later extended
to the noisy case \cite{LuVaswani12}, where a regularized modified
basis pursuit denoising method was introduced and a computable
bound on the reconstruction error was obtained. The weighted
$\ell_1$-minimization approach was also studied in
\cite{KhajehnejadXu09} by assuming a probabilistic support model
which assigns a probability of being zero or nonzero to each entry
of the sparse signal. The choice of the weights and the associated
exact recovery conditions were investigated. In
\cite{MiossoBorries09}, a modified iterative reweighted method
which incorporates the prior support information was proposed.
Similar to \cite{VaswaniLu10}, those weights corresponding to the
\emph{a priori} known support are assigned a very small value
(close to zero), whereas other weights are recursively updated
like conventional iterative reweighted methods. The problem of
support knowledge-aided sparse signal recovery was also studied in
a time-varying compressed sensing framework, e.g.
\cite{ZinielSchniter13,FangShen14dsp}, where the temporal support
correlation was modeled by a Markov chain \cite{ZinielSchniter13}
or a pattern-coupled structure \cite{FangShen14dsp}. Nevertheless,
both works need to specify the inherent support correlation
structure between two consecutive time steps, whereas our work
here deals with a more general scenario involving no particular
correlation structure between the prior support and the current
support.

%becomes more inconsistent with the groundtruth
%information becomes inconsistent with the groundtruth

It has been observed
\cite{Vaswani10,VaswaniLu10,KhajehnejadXu09,MiossoBorries09} that
the sparse recovery performance can be significantly improved
through exploiting the prior support knowledge. Nevertheless, this
improvement can only be achieved when the prior knowledge is
fairly accurate. Existing methods, e.g.
\cite{Vaswani10,VaswaniLu10,KhajehnejadXu09,MiossoBorries09}, may
suffer from severe recovery performance degradation or even
recovery failure in the presence of inaccurate prior knowledge. In
practice, however, signal support estimation inevitably incurs
errors, and in some cases, due to the support variation across
time, the prior knowledge may contain a considerable amount of
errors. In this paper, we first introduce a modified two-layer
Gaussian-inverse Gamma prior model, where an individual parameter
$b_i$, instead of a common parameter $b$ used in the conventional
framework, is employed to characterize each sparsity-controlling
hyperparameter $\alpha_i$. Through assigning different values to
the parameters $\{b_i\}$, our new prior model has the flexibility
to place non-sparsity-encouraging priors to those coefficients
that are believed in the support set. Nevertheless, the above
modified two-layer hierarchical model does not have a mechanism to
learn the true support from the partly erroneous knowledge. To
address this issue, we propose an improved hierarchical prior
model which constitutes a three-layer hierarchical form, where a
new layer is proposed in addition to the above modified two-layer
hierarchical model. The new layer places a prior on the parameters
$\{b_i\}$. Such an approach is capable of distinguishing the true
support from erroneous support through learning the values of
$\{b_i\}$. By resorting to the variational inference methodology,
we develop a new sparse Bayesian learning method which has the
ability to learn the true support from the erroneous information.

The rest of the paper is organized as follows. In Section
\ref{sec:model}, we introduce a modified two-layer
Gaussian-inverse Gamma hierarchical prior model and an improved
three-layer hierarchical prior model which enables to learn the
true support from partly erroneous knowledge. Variational Bayesian
methods are developed in Section \ref{sec:inference}. Simulation
results are provided in Section \ref{sec:simulation}, followed by
concluding remarks in Section \ref{sec:conclusion}.

\section{Hierarchical Prior Model} \label{sec:model}
We consider the problem of recovering a sparse signal
$\boldsymbol{x}\in\mathbb{R}^{n}$ from noise-corrupted
measurements
\begin{align}
\boldsymbol{y}=\boldsymbol{A}\boldsymbol{x}+\boldsymbol{w}
\end{align}
where $\boldsymbol{A}\in\mathbb{R}^{m\times n}$ ($m<n$) is the
measurement matrix, and $\boldsymbol{w}$ is the additive
multivariate Gaussian noise with zero mean and covariance matrix
$\sigma^2\boldsymbol{I}$. Suppose we have partial but partly
erroneous knowledge of the support of the sparse signal
$\boldsymbol{x}$. The prior knowledge $P$ can be divided into two
parts: $P=S\cup E$, where $S$ denotes the subset containing
correct information and $E$ denotes the error subset. If we let
$T$ denote the true support of $\boldsymbol{x}$ and $T^c$ denote
the complement of the set $T$, i.e. $T\cup T^c=\{1,2,\ldots,n\}$,
then we have $S\subset T$, and $E\subset T^c$. Note that the only
prior information we have is $P$. The partition of $S$ and $E$ is
unknown to us.

The prior support information can certainly be utilized to improve
signal recovery. In the following, based on the conventional
sparse Bayesian learning (SBL) framework, we first introduce a
modified SBL hierarchical prior model which has the flexibility to
place non-sparsity-encouraging priors to those coefficients that
are believed in the support set. Furthermore, we propose an
improved three-layer hierarchical prior model which enables to
learn the true support from the erroneous information and thus
exploits the prior support information in a more constructive
manner. To facilitate discussions and comparisons, we first
provide a brief overview of the conventional SBL hierarchical
model.

%Such a modified prior model seamlessly integrates the prior
%support information into the sparse Bayesian learning framework.

%address this issue, in this paper, we develop a sparse signal
%recovery algorithm which has the ability to distinguish the
%correct support from erroneous information and thus can exploit
%the prior support information in a more constructive way. To this
%objective, we will propose a new hierarchical sparse Bayesian
%learning (SBL) model which allows to learn the correct information
%from the partly erroneous knowledge.

%Such a problem has been studied in several parallel and
%independent works. Nevertheless, these works assume a fairly
%accurate knowledge of the true support, that is, the error set $E$
%is empty or negligible. When the knowledge becomes inaccurate,
%existing methods incur a considerable performance degradation or
%even recovery failure.

%The prior knowledge consists of a correct part and an incorrect
%part.
%overcome the barrier of incorrect prior knowledge

%Such an automatic relevance determination mechanism and the
%resulting sparse solution not only effectively avoid the
%overfitting problem, but also render superior regression and
%classification accuracy.

%to automatically remove irrelevant basis vectors and retain only a
%few `relevant' vectors to explain the observed data.

\subsection{Overview of Conventional SBL}
Sparse Bayesian learning was originally developed by Tipping in
\cite{Tipping01} to address regression and classification
problems. Later on in \cite{Wipf06,JiXue08,WipfRao07,ZhangRao11},
sparse Bayesian learning was adapted to solve the sparse recovery
problem and obtained superior performance for sparse signal
recovery in a series of experiments. In the conventional sparse
Bayesian learning framework, a two-layer hierarchical prior model
was proposed to promote the sparsity of the solution. In the first
layer, $\boldsymbol{x}$ is assigned a Gaussian prior distribution
\begin{align}
p(\boldsymbol{x}|\boldsymbol{\alpha})=\prod_{i=1}^n
p(x_i|\alpha_i) \label{hm-1}
\end{align}
where $p(x_i|\alpha_i)=\mathcal{N}(x_i|0,\alpha_i^{-1})$, and
$\boldsymbol{\alpha}\triangleq\{\alpha_i\}$, the inverse variance
(precision) of the Gaussian distribution, are non-negative
sparsity-controlling hyperparameters. The second layer specifies
Gamma distributions as hyperpriors over the hyperparameters
$\{\alpha_i\}$, i.e.
\begin{align}
p(\boldsymbol{\alpha})=\prod_{i=1}^n\text{Gamma}(\alpha_i|a,b)=\prod_{i=1}^n\Gamma(a)^{-1}b^{a}\alpha_{i}^{a-1}e^{-b\alpha_{i}}
\label{hm-2}
\end{align}
where $\Gamma(a)=\int_{0}^{\infty}t^{a-1}e^{-t}dt$ is the Gamma
function, the parameters $a$ and $b$ used to characterize the
Gamma distribution are chosen to be very small values, e.g.
$10^{-4}$, in order to provide non-informative/uniform (over a
logarithmic scale) hyperpriors over $\{\alpha_i\}$. As discussed
in \cite{Tipping01}, a broad hyperprior allows the posterior mean
of $\alpha_i$ to become arbitrarily large. As a consequence, the
associated coefficient $x_i$ will be driven to zero, thus yielding
a sparse solution. This mechanism is also referred to as the
``automatic relevance determination'' mechanism which tends to
switch off most of the coefficients that are deemed to be
irrelevant, and only keep very few relevant coefficients to
explain the data.

%\footnote{Although in \cite{Tipping01}, the parameter $a$, instead
%of $a-1$, is chosen to be a small value, }

\begin{figure*}[!t]
 \centering
\subfigure[A modified two-layer hierarchical prior
model]{\includegraphics[width=6cm]{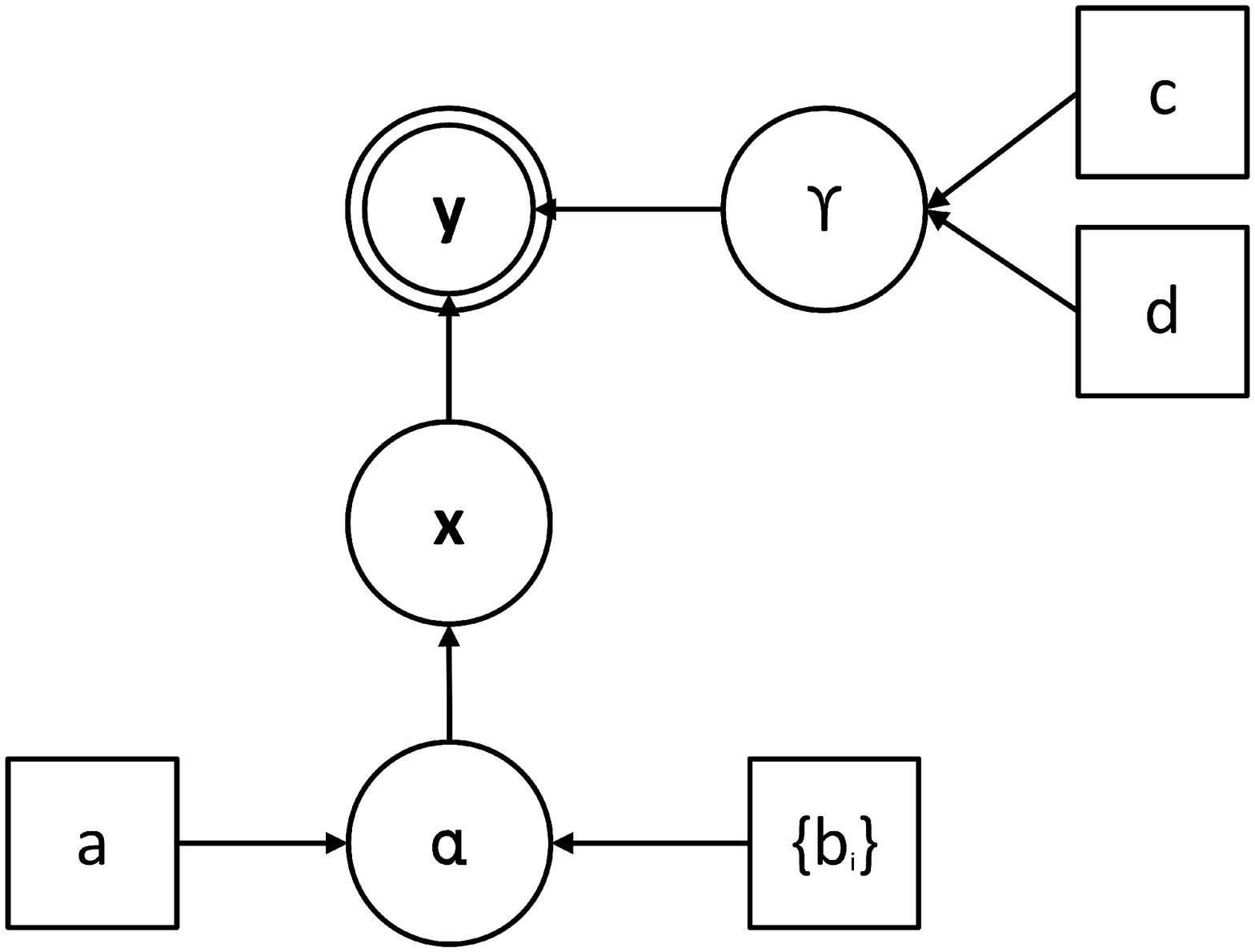}}
 \hfil
\subfigure[A three-layer hierarchical prior
model.]{\includegraphics[width=6cm]{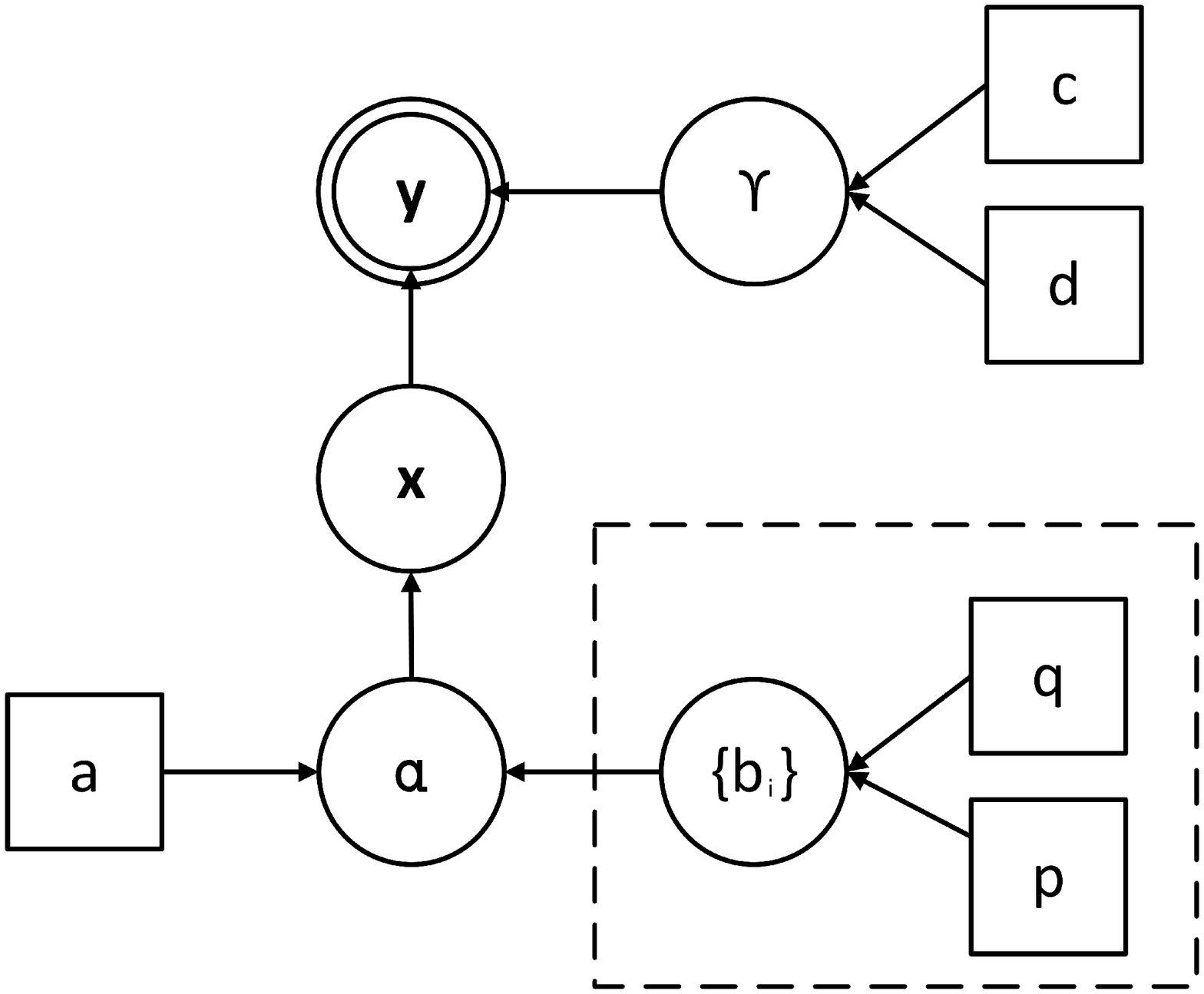}}
  \caption{Hierarchical models for support knowledge-aided sparse Bayesian learning.}
   \label{fig10}
\end{figure*}

\subsection{Modified Two-Layer Hierarchical Model}
When the value of the parameter $b$ is relatively large, e.g.
$b=0.5$, it can be readily observed from (\ref{hm-2}) that the
hyperpriors are no longer uniform and now they encourage small
values of $\{\alpha_i\}$. In this case, an arbitrarily large value
of the posterior mean of $\alpha_i$ is prohibited. As a result,
the two-layer hierarchical model no longer results in a
sparsity-encouraging prior and therefore does not necessarily lead
to a sparse solution. This fact inspires us to develop a new way
to incorporate the prior support information into the sparse
Bayesian learning framework. Specifically, instead of using a
common parameter $b$ for all sparsity-controlling hyperparameters
$\{\alpha_i\}$, we hereby employ an individual parameter $b_i$ for
each hyperparameter $\alpha_i$, i.e.
\begin{align}
p(\boldsymbol{\alpha})=\prod_{i=1}^n
\text{Gamma}(\alpha_i|a,b_i)=\prod_{i=1}^n\Gamma(a)^{-1}b_i^{a}\alpha_{i}^{a-1}e^{-b_i\alpha_{i}}
\end{align}
Such a formulation allows us to assign different priors to
different coefficients. If the partial knowledge of the signal's
support, $P$, is available, then the associated parameters of
$\{b_i\}$ can be set to a relatively large value, say $0.5$, in
order to place a non-sparsity-encouraging prior on the
corresponding coefficients, whereas the rest parameters of
$\{b_i\}$ are still assigned a small value, say $10^{-4}$, to
encourage sparse coefficients, that is,
\begin{equation}
b_i=\begin{cases}0.5 & i\in P \\ 10^{-4} & i\in P^c
\end{cases} \label{b-value}
\end{equation}
where $P^c$ denotes the complement of $P$, i.e. $P\cup
P^c=\{1,2\ldots,n\}$. The above modified hierarchical model
seamlessly integrates the prior support information into the
sparse Bayesian learning framework. Nevertheless, the modified
two-layer hierarchical model which assigns fixed values to
$\{b_i\}$ still lacks the flexibility to learn and adapt to the
true situation. When the prior information contains a considerable
portion of errors, this approach may suffer from significant
performance loss and even recovery failure.

%Based on this prior model, a variational Bayesian algorithm can be
%readily developed (details are provided in), and demonstrates
%superior recovery performance given that the partial knowledge on
%the support of the sparse signal is fairly accurate.

%i.e.
%\begin{align}
%b_i=10^{-4} \qquad \forall i\in P^c \label{hm-3-a}
%\end{align}

%\begin{align}
%\text{Gamma}(b_i|p,q)=\Gamma(p)^{-1}q^{p} b_{i}^{p-1}e^{-q b_{i}}
%\qquad \forall i\in P \label{hm-3}
%\end{align}

\subsection{Proposed Three-Layer Hierarchical Model}
To address the above issue, we partition the parameters $\{b_i\}$
into two subsets: $\{b_i, \forall i\in P\}$, and $\{b_i, \forall
i\in P^c\}$. For $\{b_i, \forall i\in P^c\}$, the parameters are
still considered to be deterministic and assigned a very small
value, e.g. $10^{-4}$; while for $\{b_i, \forall i\in P\}$,
instead of assigning a fixed large value, we model them as random
parameters and place hyperpriors over these parameters. Since
$\{b_i, \forall i\in P\}$ are positive values, suitable priors
over $\{b_i, \forall i\in P\}$ are also Gamma distributions. In
summary, we have
\begin{equation}
p(b_i)=\begin{cases}\text{Gamma}(b_i|p,q)=\Gamma(p)^{-1}q^{p}
b_{i}^{p-1}e^{-q b_{i}} & i\in P \\ \delta(b_i-10^{-4}) & i\in P^c
\end{cases} \label{hm-3}
\end{equation}
where $\delta(\cdot)$ denotes the Dirac delta function, $p$ and
$q$ are parameters characterizing the Gamma distribution and their
choice will be discussed in Section \ref{sec:inference}. We see
that the proposed model constitutes a three-layer hierarchical
form which allows to learn the parameters $\{b_i, \forall i\in
P\}$ in an automatic manner from the data. The proposed algorithm
based on this prior model therefore has the ability to identify
the correct support from erroneous information.

%The conventional sparse Bayesian learning hierarchical model con
%suppress large values

%The development reveals how the prior support information is
%seamlessly incorporated into the developed algorithm.

%Note that in the previous section, a modified two-layer
%hierarchical model and a three-layer hierarchical model were
%introduced. In this section, we will first consider the modified
%two-layer hierarchical model where the parameters $\{b_i\}$ are
%assigned fixed values. We then focus on the Bayesian inference for
%the three-layer model which provides flexibility in learning the
%correct support from partly erroneous information.

\section{Variational Bayesian Inference} \label{sec:inference}
We now proceed to perform variational Bayesian inference for the
proposed hierarchical models. Throughout this paper, the noise
variance $\sigma^2$ is assumed unknown, and needs to be estimated
along with other parameters. For notational convenience, define
\begin{align}
\gamma\triangleq\sigma^{-2} \nonumber
\end{align}
Following the conventional sparse Bayesian learning framework
\cite{Tipping01}, we place a Gamma hyperprior over $\gamma$:
\begin{align}
p(\gamma)=\text{Gamma}(\gamma|c,d)=\Gamma(c)^{-1}d^{c}\gamma^{c-1}e^{-d\gamma}
\label{gamma-prior}
\end{align}
where the parameters $c$ and $d$ are set to small values, e.g.
$c=d=10^{-4}$. Before proceeding, we firstly provide a brief
review of the variational Bayesian methodology.

\subsection{Review of The Variational Bayesian Methodology}
In a probabilistic model, let $\boldsymbol{y}$ and
$\boldsymbol{\theta}$ denote the observed data and the hidden
variables, respectively. It is straightforward to show that the
marginal probability of the observed data can be decomposed into
two terms
\begin{align}
\ln p(\boldsymbol{y})=L(q)+\text{KL}(q|| p)
\label{variational-decomposition}
\end{align}
where
\begin{align}
L(q)=\int q(\boldsymbol{\theta})\ln
\frac{p(\boldsymbol{y},\boldsymbol{\theta})}{q(\boldsymbol{\theta})}d\boldsymbol{\theta}
\end{align}
and
\begin{align}
\text{KL}(q|| p)=-\int q(\boldsymbol{\theta})\ln
\frac{p(\boldsymbol{\theta}|\boldsymbol{y})}{q(\boldsymbol{\theta})}d\boldsymbol{\theta}
\end{align}
where $q(\boldsymbol{\theta})$ is any probability density
function, $\text{KL}(q|| p)$ is the Kullback-Leibler divergence
between $p(\boldsymbol{\theta}|\boldsymbol{y})$ and
$q(\boldsymbol{\theta})$. Since $\text{KL}(q|| p)\geq 0$, it
follows that $L(q)$ is a rigorous lower bound on $\ln
p(\boldsymbol{y})$. Moreover, notice that the left hand side of
(\ref{variational-decomposition}) is independent of
$q(\boldsymbol{\theta})$. Therefore maximizing $L(q)$ is
equivalent to minimizing $\text{KL}(q|| p)$, and thus the
posterior distribution $p(\boldsymbol{\theta}|\boldsymbol{y})$ can
be approximated by $q(\boldsymbol{\theta})$ through maximizing
$L(q)$.

%by assuming an appropriate $q(\boldsymbol{\theta})$

The significance of the above transformation is that it
circumvents the difficulty of computing the posterior probability
$p(\boldsymbol{\theta}|\boldsymbol{y})$ directly (which is usually
computationally intractable). For a suitable choice for the
distribution $q(\boldsymbol{\theta})$, the quantity $L(q)$ may be
more amiable to compute. Specifically, we could assume some
specific parameterized functional form for $q$ and then maximize
$L(q)$ with respect to the parameters of the distribution. A
particular form of $q(\boldsymbol{\theta})$ that has been widely
used with great success is the factorized form over the component
variables $\{\theta_i\}$ in $\boldsymbol{\theta}$, i.e.
\begin{align}
q(\boldsymbol{\theta})=\prod_i q_i(\theta_i) \label{factorization}
\end{align}
We therefore can compute the posterior distribution approximation
by finding $q(\boldsymbol{\theta})$ of the form
(\ref{factorization}) that maximizes the lower bound $L(q)$. The
maximization can be conducted in an alternating fashion for each
latent variable, which leads to \cite{TzikasLikas08}
\begin{align}
q_i(\theta_i)=\frac{\exp(\langle\ln
p(\boldsymbol{y},\boldsymbol{\theta})\rangle_{k\neq
i})}{\int\exp(\langle\ln
p(\boldsymbol{y},\boldsymbol{\theta})\rangle_{k\neq i})d\theta_i}
\end{align}
where $\langle\cdot\rangle_{k\neq i}$ denotes an expectation with
respect to the distributions $q(\theta_i)$ for all $k\neq i$.

\subsection{Bayesian Inference for Modified Two-Layer Model} \label{sec:inference-NSL}
Let
$\boldsymbol{\theta}\triangleq\{\boldsymbol{x},\boldsymbol{\alpha},
\boldsymbol{\gamma}\}$ denote all hidden variables. We assume
posterior independence among the variables $\boldsymbol{x}$,
$\boldsymbol{\alpha}$, and $\gamma$, i.e.
\begin{align}
p(\boldsymbol{\theta}|\boldsymbol{y})\approx &
q(\boldsymbol{x},\boldsymbol{\alpha},\gamma)\nonumber\\
=&
q_x(\boldsymbol{x})q_{\alpha}(\boldsymbol{\alpha})q_{\gamma}(\gamma)
\end{align}
With this mean field approximation, the posterior distribution of
each hidden variable can be computed by maximizing $L(q)$ while
keeping other variables fixed using their most recent
distributions, which gives
\begin{align}
\ln q_x(\boldsymbol{x})=&\langle\ln
p(\boldsymbol{y},\boldsymbol{x},\boldsymbol{\alpha},\gamma)\rangle_{q_{\alpha}(\boldsymbol{\alpha})
q_{\gamma}(\gamma)} + \text{constant} \nonumber\\
\ln q_{\alpha}(\boldsymbol{\alpha})=&\langle\ln
p(\boldsymbol{y},\boldsymbol{x},\boldsymbol{\alpha},\gamma)\rangle_{q_x(\boldsymbol{x})
q_{\gamma}(\gamma)} + \text{constant} \nonumber\\
\ln q_{\gamma}(\gamma)=&\langle\ln
p(\boldsymbol{y},\boldsymbol{x},\boldsymbol{\alpha},\gamma)\rangle_{q_x(\boldsymbol{x})
q_{\alpha}(\boldsymbol{\alpha})} + \text{constant} \nonumber
\end{align}
where $\langle\rangle_{q(\cdot)}$ denotes the expectation with
respect to (w.r.t.) the distribution $q(\cdot)$. In summary, the
posterior distribution approximations are computed in an
alternating fashion for each hidden variable, with other variables
fixed. Details of this Bayesian inference scheme are provided
below.

\textbf{\emph{1). Update of $q_x(\boldsymbol{x})$}}: Keeping only
the terms that depend on $\boldsymbol{x}$, the variational
optimization of $q_{x}(\boldsymbol{x})$ yields
\begin{align}
\ln q_x(\boldsymbol{x})\propto & \langle\ln
p(\boldsymbol{y}|\boldsymbol{x},\gamma)+\ln
p(\boldsymbol{x}|\boldsymbol{\alpha})
\rangle_{q_{\alpha}(\boldsymbol{\alpha})q_{\gamma}(\gamma)}
\nonumber\\
\propto &
-\frac{\langle\gamma\rangle}{2}(\boldsymbol{y}-\boldsymbol{A}\boldsymbol{x})^T(\boldsymbol{y}-\boldsymbol{A}\boldsymbol{x})
-\frac{1}{2}\boldsymbol{x}^T\langle
\boldsymbol{D}\rangle\boldsymbol{x}
\end{align}
where $\langle\gamma\rangle$ denotes the expectation w.r.t.
$q_{\gamma}(\gamma)$, and $\langle
\boldsymbol{D}\rangle\triangleq\text{diag}(\langle
\alpha_1\rangle, \ldots, \langle \alpha_n\rangle)$, in which
$\langle \alpha_i\rangle$ represents the expectation w.r.t.
$q_{\alpha}(\alpha)$. It can be readily verified that
$q_x(\boldsymbol{x})$ follows a Gaussian distribution with its
mean $\boldsymbol{\mu}$ and covariance matrix $\boldsymbol{\Phi}$
given respectively as
\begin{align}
\boldsymbol{\mu}=&\langle\gamma\rangle\boldsymbol{\Phi}\boldsymbol{A}^T\boldsymbol{y}
\nonumber\\
\boldsymbol{\Phi}=&\left(\langle\gamma\rangle\boldsymbol{A}^T\boldsymbol{A}+\langle\boldsymbol{D}\rangle\right)^{-1}
\label{x-update}
\end{align}

\textbf{\emph{2). Update of $q_{\alpha}(\boldsymbol{\alpha})$}}:
Similarly, the approximate posterior
$q_{\alpha}(\boldsymbol{\alpha})$ can be obtained by computing
\begin{align}
\ln q_{\alpha}(\boldsymbol{\alpha}) \propto & \langle\ln
p(\boldsymbol{x}|\boldsymbol{\alpha})+\ln
p(\boldsymbol{\alpha}|a,\boldsymbol{b}) \rangle_{q_x(\boldsymbol{x})} \nonumber\\
\propto & \sum_{i=1}^n\left\{
\left(a-\frac{1}{2}\right)\ln\alpha_i-\left(b_i+\frac{\langle
x_i^2\rangle}{2}\right)\alpha_i \right\}
\end{align}
where $\langle x_i^2\rangle$ denotes the expectation w.r.t.
$q_x(\boldsymbol{x})$. Thus $\boldsymbol{\alpha}$ has a form of a
product of Gamma distributions
\begin{align}
q_{\alpha}(\boldsymbol{\alpha})=\prod_{i=1}^n
\text{Gamma}(\alpha_i; \tilde{a}, \tilde{b}_i)
\end{align}
in which the parameters $\tilde{a}$ and $\tilde{b}_i$ are
respectively given as
\begin{align}
\tilde{a}=a+\frac{1}{2} \qquad \tilde{b}_i=b+\frac{1}{2}\langle
x_i^2\rangle \label{alpha-update}
\end{align}

\textbf{\emph{3). Update of $q_{\gamma}(\boldsymbol{\gamma})$}}:
The approximate posterior distribution
$q_{\gamma}(\boldsymbol{\gamma})$ can be computed as
\begin{align}
\ln q_{\gamma}(\boldsymbol{\gamma}) \propto & \langle\ln
p(\boldsymbol{y}|\boldsymbol{x},\gamma)+\ln
p(\gamma|c,d)\rangle_{q_x(\boldsymbol{x})} \nonumber\\
\propto & \left(\frac{m}{2}+c\right)\ln\gamma
\nonumber\\
&-\left(\frac{1}{2}\langle(\boldsymbol{y}-\boldsymbol{A}\boldsymbol{x})^T
(\boldsymbol{y}-\boldsymbol{A}\boldsymbol{x})\rangle_{q_x(\boldsymbol{x})}+d\right)\gamma
\end{align}
We can easily see that $q(\gamma)$ follows a Gamma distribution
\begin{align}
q(\gamma)=\text{Gamma}(\gamma|\tilde{c},\tilde{d})
\label{gamma-posterior}
\end{align}
with the parameters $\tilde{c}$ and $\tilde{d}$ given respectively
by
\begin{align}
\tilde{c}=&\frac{m}{2}+c \nonumber\\
\tilde{d}=&d+\frac{1}{2}\langle(\boldsymbol{y}-\boldsymbol{A}\boldsymbol{x})^T
(\boldsymbol{y}-\boldsymbol{A}\boldsymbol{x})\rangle_{q_x(\boldsymbol{x})}
\label{gamma-update}
\end{align}
where
\begin{align}
\langle(\boldsymbol{y}-\boldsymbol{A}\boldsymbol{x})^T
(\boldsymbol{y}-\boldsymbol{A}\boldsymbol{x})\rangle_{q_x(\boldsymbol{x})}=
\|\boldsymbol{y}-\boldsymbol{A}\boldsymbol{\mu}\|_2^2+
\text{tr}\left\{\boldsymbol{A}^T\boldsymbol{A}\boldsymbol{\Phi}\right\}
\nonumber
\end{align}

In summary, the variational Bayesian inference involves updates of
the approximate posterior distributions for hidden variables
$\boldsymbol{x}$, $\boldsymbol{\alpha}$, and $\gamma$. Some of the
expectations and moments used during the update are summarized as
\begin{align}
\langle \alpha_i\rangle=&\frac{\tilde{a}}{\tilde{b}_i} \quad
\langle\gamma\rangle=\frac{\tilde{c}}{\tilde{d}} \nonumber\\
\langle x_i^2\rangle=&\mu_{i}^2+\phi_{i,i}  \nonumber
\end{align}
where $\mu_{i}$ denotes the $i$th element of $\boldsymbol{\mu}$,
and $\phi_{i,i}$ denotes the $i$th diagonal element of
$\boldsymbol{\Phi}$. For clarity, we summarize our algorithm as
follows.

%\begin{table}[h]
\begin{center}
\textbf{Support Aided-SBL with No Support Learning}
\end{center}
%\begin{tabular}{p{8cm}}
\vspace{0cm} \noindent
\begin{tabular}{lp{7.7cm}}
\hline 1.& Given the current approximate posterior distributions
$q_{\alpha}(\boldsymbol{\alpha})$ and $q_{\gamma}(\gamma)$, update
the posterior distribution $q_{x}(\boldsymbol{x})$ according
to (\ref{x-update}).\\
2.& Given $q_{\gamma}(\gamma)$ and $q_{x}(\boldsymbol{x})$, update
$q_{\alpha}(\boldsymbol{\alpha})$ according to (\ref{alpha-update}).\\
3.& Given $q_{\alpha}(\boldsymbol{\alpha})$ and
$q_{x}(\boldsymbol{x})$, update $q_{\gamma}(\gamma)$ according to
(\ref{gamma-update}).\\
4.& Continue the above iterations until
$\|\boldsymbol{\mu}^{(t+1)}-\boldsymbol{\mu}^{(t)}\|_2\leq\epsilon$,
where $\epsilon$ is a prescribed tolerance value.\\
\hline
\end{tabular}
%\end{table}

\vspace{0.3cm}

The update for $\langle \alpha_i\rangle$ can be written as
\begin{align}
\langle \alpha_i\rangle=\frac{1+2a}{\langle x_i^2\rangle+2b_i}
\label{alphai-update}
\end{align}
We observe that this update rule is similar to that of the
conventional SBL, except that the common parameter $b$ is now
replaced by a set of individual parameters $\{b_i\}$. When $b_i$
is very small, e.g. $10^{-4}$, the update for
$\langle\alpha_i\rangle$ is exactly the same as the conventional
rule. Hence the Bayesian Occam's razor which contributes to the
success of the conventional SBL also works here for those
coefficients whose corresponding parameters $\{b_i\}$ are small.
To see this, note that when computing the posterior mean and
covariance matrix, a large $\langle\alpha_i\rangle$ tends to
suppress the values of the corresponding components
$\{\mu_i,\phi_{i,i}\}$ (c.f. (\ref{x-update})), which in turn
leads to a larger $\langle\alpha_i\rangle$ (c.f.
(\ref{alphai-update})). This negative feedback mechanism keeps
decreasing most of the coefficients until they become negligible,
while leaving only a few prominent nonzero entries survived to
explain the data. On the other hand, when $b_i$ is set to be
relatively large, e.g. $0.5$, from (\ref{alphai-update}) we know
that $\langle\alpha_i\rangle$ cannot be arbitrarily large and thus
the automatic relevance determination mechanism is disabled. In
other words, for those coefficients whose corresponding parameters
$\{b_i\}$ are large, the priors assigned to these coefficients are
no longer sparsity-encouraging. Hence the prior support
information is seamlessly incorporated into the proposed algorithm
by assigning different values to $\{b_i\}$.

%in which case $\alpha_i$ has the tendency to become very large,
%and as a consequence, the associated coefficient $x_i$ will
%decrease to zero.

%Considering that the hierarchical model is more complex than the
%previous two-layer model, we use

\subsection{Bayesian Inference for Proposed Three-Layer Model}
\label{sec:inference-SL} When the prior support set $P$ contains a
considerable portion of errors, the above proposed algorithm may
incur a significant performance loss because those parameters
$\{b_i, \forall i\in E\}$ supposed to be very small are assigned
large values. Instead, the proposed three-layer hierarchical model
provides a flexible framework for adaptively learning values of
$\{b_i, \forall i\in P\}$.

%In this section, we resort to the variational inference
%methodology to circumvent the computation of the exact posterior
%distribution of hidden variables.

%Conducting Bayesian inference by exploiting the EM formulation for
%the three-stage hierarchical model, however, is complex because it
%requires to compute the posterior distribution of hidden
%variables, which may not admit an analytical form.

%in an alternating fashion for each hidden variable.

%Specifically, let
%$\boldsymbol{z}=\{\boldsymbol{x},\boldsymbol{\alpha},\gamma,\boldsymbol{\bar{b}}\}$
%be the vector of all hidden variables.

Let $\boldsymbol{\theta}\triangleq\{\boldsymbol{x},
\boldsymbol{\alpha}, \gamma, \boldsymbol{\bar{b}}\}$, where
$\boldsymbol{\bar{b}}\triangleq \{b_i, \forall i\in P\}$ are
hidden variables as well since they are assigned hyperpriors and
need to be learned. We assume posterior independence among the
hidden variables $\boldsymbol{x}$, $\boldsymbol{\alpha}$,
$\gamma$, and $\boldsymbol{\bar{b}}$, i.e.
\begin{align}
p(\boldsymbol{x},\boldsymbol{\alpha},\gamma,\boldsymbol{\bar{b}}|\boldsymbol{y})\approx
&
q(\boldsymbol{x},\boldsymbol{\alpha},\gamma,\boldsymbol{\bar{b}})\nonumber\\
=&
q_x(\boldsymbol{x})q_{\alpha}(\boldsymbol{\alpha})q_{\gamma}(\gamma)q_{\bar{b}}(\boldsymbol{\bar{b}})
\end{align}
With this mean field approximation, the posterior distribution of
each hidden variable can be computed by minimizing the
Kullback-Leibler (KL) divergence while keeping other variables
fixed using their most recent distributions, which gives
\begin{align}
\ln q_x(\boldsymbol{x})=&\langle\ln
p(\boldsymbol{y},\boldsymbol{x},\boldsymbol{\alpha},\gamma,\boldsymbol{\bar{b}})\rangle_{q_{\alpha}(\boldsymbol{\alpha})
q_{\gamma}(\gamma)q_{\bar{b}}(\boldsymbol{\bar{b}})} + \text{constant} \nonumber\\
\ln q_{\alpha}(\boldsymbol{\alpha})=&\langle\ln
p(\boldsymbol{y},\boldsymbol{x},\boldsymbol{\alpha},\gamma,\boldsymbol{\bar{b}})\rangle_{q_x(\boldsymbol{x})
q_{\gamma}(\gamma)q_{\bar{b}}(\boldsymbol{\bar{b}})} + \text{constant} \nonumber\\
\ln q_{\gamma}(\gamma)=&\langle\ln
p(\boldsymbol{y},\boldsymbol{x},\boldsymbol{\alpha},\gamma,\boldsymbol{\bar{b}})\rangle_{q_x(\boldsymbol{x})
q_{\alpha}(\boldsymbol{\alpha})q_{\bar{b}}(\boldsymbol{\bar{b}})} + \text{constant} \nonumber\\
\ln q_{\bar{b}}(\boldsymbol{\bar{b}})=&\langle\ln
p(\boldsymbol{y},\boldsymbol{x},\boldsymbol{\alpha},\gamma,\boldsymbol{\bar{b}})\rangle_{q_x(\boldsymbol{x})
q_{\alpha}(\boldsymbol{\alpha})q_{\gamma}(\gamma)} +
\text{constant} \nonumber
\end{align}
Details of this Bayesian inference scheme are provided below.

%where $\langle\rangle_{q(\cdot)}$ denotes the expectation with
%respect to the distribution $q(\cdot)$. In summary, the posterior
%distribution approximations are computed in an alternating fashion
%for each hidden variable, with other variables fixed.

\textbf{\emph{1). Update of $q_x(\boldsymbol{x})$}}: The
variational optimization of $q_x(\boldsymbol{x})$ can be
calculated as follows by ignoring the terms that are independent
of $\boldsymbol{x}$:
\begin{align}
\ln q(\boldsymbol{x})\propto&\langle\ln
p(\boldsymbol{y}|\boldsymbol{x},\gamma)+\ln
p(\boldsymbol{x}|\boldsymbol{\alpha})\rangle_{q_{\alpha}(\boldsymbol{\alpha})q_{\gamma}(\gamma)} \nonumber\\
\propto&
-\frac{\langle\gamma\rangle}{2}(\boldsymbol{y}-\boldsymbol{A}\boldsymbol{x})^T(\boldsymbol{y}-\boldsymbol{A}\boldsymbol{x})
-\frac{1}{2}\boldsymbol{x}^T\langle
\boldsymbol{D}\rangle\boldsymbol{x}
\end{align}
We can easily verify that $q(\boldsymbol{x})$ follows a Gaussian
distribution with its mean $\boldsymbol{\mu}$ and covariance
matrix $\boldsymbol{\Phi}$ given respectively as
\begin{align}
\boldsymbol{\mu}=&\langle\gamma\rangle\boldsymbol{\Phi}\boldsymbol{A}^T\boldsymbol{y}
\nonumber\\
\boldsymbol{\Phi}=&\left(\langle\gamma\rangle\boldsymbol{A}^T\boldsymbol{A}+\langle\boldsymbol{D}\rangle\right)^{-1}
\label{x-update-2}
\end{align}

\textbf{\emph{2). Update of $q_{\alpha}(\boldsymbol{\alpha})$}}:
Similarly, the approximate posterior
$q_{\alpha}(\boldsymbol{\alpha})$ can be computed as
\begin{align}
\ln q_{\alpha}(\boldsymbol{\alpha}) \propto& \langle\ln
p(\boldsymbol{x}|\boldsymbol{\alpha})+\ln
p(\boldsymbol{\alpha}|a,\boldsymbol{b})
\rangle_{q_x(\boldsymbol{x})q_{\bar{b}}(\boldsymbol{\bar{b}})} \nonumber\\
=& \sum_i^n\left\langle (a-0.5)\ln\alpha_i-(0.5 x_i^2+b_i)\alpha_i
\right\rangle_{q_x(\boldsymbol{x})q_{\bar{b}}(\boldsymbol{\bar{b}})}
\nonumber\\
\stackrel{(a)}{=}& \sum_{i\in P}
\left\{(a+0.5)\ln\alpha_i-(\langle
b_i\rangle+0.5\langle x_i^2\rangle)\alpha_i\right\} \nonumber\\
& + \sum_{i\in P^c} \left\{(a+0.5)\ln\alpha_i-( b_i+0.5\langle
x_i^2\rangle)\alpha_i\right\}
\end{align}
where in $(a)$, the terms inside the summation are partitioned
into two subsets $P$ and $P^c$ because $\{b_i, i\in P^c\}$ are
deterministic parameters, while $\{b_i, i\in P\}$ are latent
variables and thus we need to perform the expectation over these
hidden variables. The posterior $q(\boldsymbol{\alpha})$ has a
form of a product of Gamma distributions
\begin{align}
q(\boldsymbol{\alpha})=\prod_{i=1}^n
\text{Gamma}(\alpha_i|\tilde{a},\tilde{b}_i)
\label{alpha-posterior}
\end{align}
with the parameters $\tilde{a}$ and $\tilde{b}_i$ given by
\begin{equation}
\tilde{a}=a+0.5 \label{tilde-a}
\end{equation}
\begin{equation}
\tilde{b}_i=\begin{cases} \langle b_i\rangle+0.5\langle
x_i^2\rangle & i\in P \\
b_i+0.5\langle x_i^2\rangle & i\in P^c \end{cases} \label{tilde-b}
\end{equation}
%in which $\langle x_i^2\rangle=\mu_i^2+\phi_{i,i}$, where $\mu_i$
%denotes the $i$th entry of $\boldsymbol{\mu}$, $\phi_{i,i}$
%denotes the $i$th diagonal element of the covariance matrix
%$\boldsymbol{\Phi}$. Also, recalling the property of the Gamma
%distribution, we have
%\begin{align}
%\langle\alpha_i\rangle=\frac{\tilde{a}+1}{\tilde{b}_i}
%\label{alpha-posterior-mean-update}
%\end{align}
%in which $\langle x_i^2\rangle=\mu_i^2+\phi_{i,i}$, $\mu_i$

%whose values are given in (\ref{hm-3-a})

\textbf{\emph{3). Update of $q_{\gamma}(\gamma)$}}: The
approximate posterior $q_{\gamma}(\gamma)$ can be computed as
\begin{align}
\ln q_{\gamma}(\gamma)\propto& \langle\ln
p(\boldsymbol{y}|\boldsymbol{x},\gamma)+\ln p(\gamma|c,d)
\rangle_{q_x(\boldsymbol{x})} \nonumber\\
\propto&\left(\frac{m}{2}+c-1\right)\ln\gamma
\nonumber\\
&-\left(\frac{1}{2}\langle(\boldsymbol{y}-\boldsymbol{A}\boldsymbol{x})^T
(\boldsymbol{y}-\boldsymbol{A}\boldsymbol{x})\rangle_{q_x(\boldsymbol{x})}+d\right)\gamma
\end{align}
It can be easily verified that $q(\gamma)$ follows a Gamma
distribution
\begin{align}
q(\gamma)=\text{Gamma}(\gamma|\tilde{c},\tilde{d})
\label{gamma-posterior}
\end{align}
where
\begin{align}
\tilde{c}=&\frac{m}{2}+c \nonumber\\
\tilde{d}=&d+\frac{1}{2}\langle(\boldsymbol{y}-\boldsymbol{A}\boldsymbol{x})^T
(\boldsymbol{y}-\boldsymbol{A}\boldsymbol{x})\rangle_{q_x(\boldsymbol{x})}
\label{tilde-d}
\end{align}
in which
\begin{align}
\langle(\boldsymbol{y}-\boldsymbol{A}\boldsymbol{x})^T
(\boldsymbol{y}-\boldsymbol{A}\boldsymbol{x})\rangle_{q_x(\boldsymbol{x})}=
\|\boldsymbol{y}-\boldsymbol{A}\boldsymbol{\mu}\|_2^2+
\text{tr}\left\{\boldsymbol{A}^T\boldsymbol{A}\boldsymbol{\Phi}\right\}
\nonumber
\end{align}
%The mean of the posterior probability of $\gamma$ can be obtained
%as
%\begin{align}
%\langle\gamma\rangle=\frac{\tilde{c}+1}{\tilde{d}}
%\end{align}

\textbf{\emph{4). Update of $q_{\bar{b}}(\boldsymbol{\bar{b}})$}}:
The variational optimization of
$q_{\bar{b}}(\boldsymbol{\bar{b}})$ yields:
\begin{align}
\ln q_{\bar{b}}(\boldsymbol{\bar{b}})\propto&\langle\ln
p(\boldsymbol{\alpha}|a,\boldsymbol{b})+\ln
p(\boldsymbol{\bar{b}}|p,q)\rangle_{q_{\alpha}(\boldsymbol{\alpha})}
\nonumber\\
\propto&\sum_{i\in P}\{-b_i\langle\alpha_i\rangle+(p-1)\ln b_i-q
b_i\}
\end{align}
from which we can readily arrive at
\begin{align}
q(\boldsymbol{\bar{b}})=\prod_{i\in P}
\text{Gamma}(b_i|p,\tilde{q}_i) \label{bar-b-posterior}
\end{align}
where
\begin{align}
\tilde{q}_i=q+\langle\alpha_i\rangle \nonumber
\end{align}
%The mean of the posterior distribution of $\{b_i, \forall i\in
%P\}$ is given by
%\begin{align}
%\langle
%b_i\rangle=\frac{p+1}{\tilde{q}_i}=\frac{p+1}{q+\langle\alpha_i\rangle}
%\label{bi-posterior-mean-update}
%\end{align}

In summary, the variational Bayesian inference consists of
successive updates of the approximate posterior distributions for
hidden variables $\boldsymbol{x}$, $\boldsymbol{\alpha}$,
$\gamma$, and $\boldsymbol{\bar{b}}$. Some of the expectations and
moments used during the update are summarized as
\begin{align}
\langle \alpha_i\rangle=\frac{\tilde{a}}{\tilde{b}_i} \quad &
\langle\gamma\rangle=\frac{\tilde{c}}{\tilde{d}} \nonumber\\
\langle x_i^2\rangle=\mu_{i}^2+\phi_{i,i} \quad & \langle
b_i\rangle=\frac{p}{\tilde{q}_i}  \nonumber
\end{align}
where $\mu_{i}$ denotes the $i$th element of $\boldsymbol{\mu}$,
and $\phi_{i,i}$ denotes the $i$th diagonal element of
$\boldsymbol{\Phi}$. We now summarize our algorithm as follows.

\begin{center}
\textbf{Partial Support Aided-SBL with Support Learning}
\end{center}
%\begin{tabular}{p{8cm}}
\vspace{0cm} \noindent
\begin{tabular}{lp{7.7cm}}
\hline 1.& Given the current approximate posterior distributions
of $q_{\alpha}(\boldsymbol{\alpha})$, $q_{\gamma}(\gamma)$ and
$q_{\bar{b}}(\boldsymbol{\bar{b}})$, update $q_x(\boldsymbol{x})$
according to (\ref{x-update-2}).\\
2.& Given $q_x(\boldsymbol{x})$, $q_{\gamma}(\gamma)$ and
$q_{\bar{b}}(\boldsymbol{\bar{b}})$, update
$q_{\alpha}(\boldsymbol{\alpha})$ according to (\ref{alpha-posterior})--(\ref{tilde-b}).\\
3.& Given $q_x(\boldsymbol{x})$,
$q_{\alpha}(\boldsymbol{\alpha})$, and
$q_{\bar{b}}(\boldsymbol{\bar{b}})$, update $q_{\gamma}(\gamma)$
according to
(\ref{gamma-posterior})--(\ref{tilde-d}).\\
4.& Given $q_x(\boldsymbol{x})$, $q_{\alpha}(\boldsymbol{\alpha})$
and $q_{\gamma}(\gamma)$, update
$q_{\bar{b}}(\boldsymbol{\bar{b}})$ according to
(\ref{bar-b-posterior}).\\
5.& Continue the above iterations until
$\|\boldsymbol{\mu}^{(t)}-\boldsymbol{\mu}^{(t-1)}\|_2\leq\epsilon$,
where $\epsilon$ is a prescribed tolerance value. Choose
$\boldsymbol{\hat{\mu}}^{(t)}$ as the estimate
of the sparse signal.\\
\hline
\end{tabular}
%\end{table}

\vspace{0.3cm}

The above proposed algorithm has the ability to adaptively learn
the values of $\{b_i,\forall i\in P\}$, and thus has the potential
to distinguish the correct support from erroneous information. To
gain insight into the algorithm, we examine the update rules for
$\langle\alpha_i\rangle$ and $\langle b_i\rangle$, i.e. the
posterior means of $\alpha_i$ and $b_i$. The update rules are
given by
\begin{equation}
\langle\alpha_i\rangle=\begin{cases} \frac{1+2a}{\langle
x_i^2\rangle+2\langle b_i\rangle} & i\in P \\
\frac{1+2a}{\langle x_i^2\rangle+2b_i} & i\in P^c
\end{cases} \label{alpha-posterior-mean-update}
\end{equation}
and
\begin{align}
\langle b_i\rangle=\frac{p}{q+\langle\alpha_i\rangle}
\label{bi-posterior-mean-update}
\end{align}
respectively. We see that the two update rules are related to each
other, with $\langle b_i\rangle$ used in updating
$\langle\alpha_i\rangle$ and $\langle\alpha_i\rangle$ used in
obtaning a new $\langle b_i\rangle$. A closer examination reveals
that $\langle\alpha_i\rangle$ and $\langle b_i\rangle$ are
inversely proportional to each other in their respective update
rules. Specifically, a smaller $\langle b_i\rangle$ results in a
larger $\langle\alpha_i\rangle$ (c.f.
(\ref{alpha-posterior-mean-update})), which in turn leads to a
smaller $\langle b_i\rangle$ (c.f.
(\ref{bi-posterior-mean-update})). This negative feedback
mechanism could keep decreasing $\langle b_i\rangle$ until it
becomes negligible, and eventually lead to an arbitrarily large
$\langle\alpha_i\rangle$. Nevertheless, the interaction between
$\langle b_i\rangle$ and $\langle\alpha_i\rangle$ could go the
other way around, i.e. a larger $\langle b_i\rangle$ results in a
smaller $\langle\alpha_i\rangle$, which in turn leads to a larger
$\langle b_i\rangle$. In this case, $\langle\alpha_i\rangle$ will
eventually converge to a finite value. The behavior of $\langle
x_i^2\rangle$ plays a key role in determining the course of the
evolution, and in turn has an impact on the dynamic behavior of
$\langle x_i^2\rangle$ itself. With a proper choice of $p$ and
$q$, if $\langle x_i^2\rangle$ becomes sufficiently small during
the iterative process, then the process could evolve towards the
$\langle\alpha_i\rangle\rightarrow\infty$ (that is, $\langle
x_i^2\rangle\rightarrow 0$) direction, otherwise the process will
converge to a finite $\langle\alpha_i\rangle$, in which case a
non-sparsity-encouraging prior is imposed on the coefficient
$x_i$. This, clearly, is a sensible strategy to learn the
parameters $\{b_i\}$. We now discuss the choice of the parameters
$p$ and $q$. To enable an efficient interaction between $\langle
b_i\rangle$ and $\langle\alpha_i\rangle$, we hope $\langle
\alpha_i\rangle$ plays a critical role in determining the value of
$\langle b_i\rangle$. To this goal, the values of $p$ and $q$
should be set sufficiently small. Our experiments suggest that
$p=q=0.1$ is a suitable choice which enables effective support
learning.

%Suppose $\langle x_i^2\rangle$ keeps decreasing during the
%iterative process, then ,

%Given that $\langle x_i^2\rangle$ tends to become smaller and
%smaller,

%considering (\ref{alpha-posterior-mean-update}), when $\langle
%b_i\rangle$ is close to zero, then $\langle\alpha_i\rangle$ is
%allowed to become arbitrarily large, in which case the automatic
%relevance determination mechanism is enabled for the coefficient
%$x_i$; when $\langle b_i\rangle$ is relatively large,
%$\langle\alpha_i\rangle$ has a finite value which cannot be very
%large, thus the prior assigned to $x_i$ is no longer
%sparsity-encouraging.

%As observed from (\ref{bi-posterior-mean-update}), $\langle
%b_i\rangle$ is inversely proportional to $\langle\alpha_i\rangle$,
%that is, the larger $\langle\alpha_i\rangle$, the smaller $\langle
%b_i\rangle$, and vice versa.

%where $\tilde{a}$ and $\tilde{b}_i$ are respectively given in
%(\ref{tilde-b}).

\section{Simulation Results} \label{sec:simulation}
We now carry out experiments to illustrate the performance of our
proposed algorithms and compare with other existing methods. The
proposed algorithm in Section \ref{sec:inference-NSL} is referred
to as the support knowledge-aided sparse Bayesian learning with no
support learning (SA-SBL-NSL), and the one in Section
\ref{sec:inference-SL} as the support knowledge-aided sparse
Bayesian learning with support learning (SA-SBL-SL), respectively.
The performance of the proposed algorithms\footnote{Matlab codes
for our algorithm are available at
http://www.junfang-uestc.net/codes/SA-SBL.rar} will be examined
using both synthetic and real data. Throughout our experiments,
the parameters $p$ and $q$ for our proposed algorithm are set
equal to $p=q=0.1$.

\begin{figure}[!t]
\centering
\includegraphics[width=8cm]{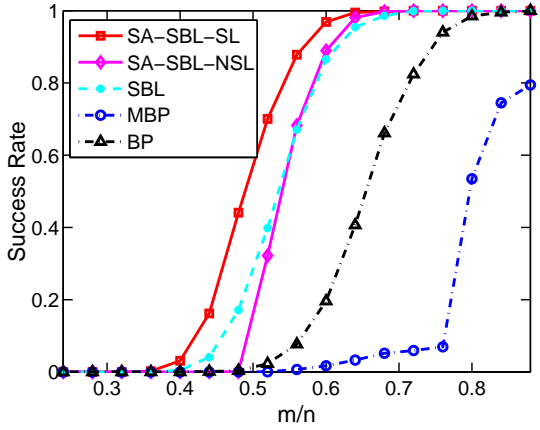}
\caption{Success rates of respective algorithms vs. the ratio
$m/n$.} \label{fig1}
\end{figure}

\begin{figure}[!t]
\centering
\includegraphics[width=8cm]{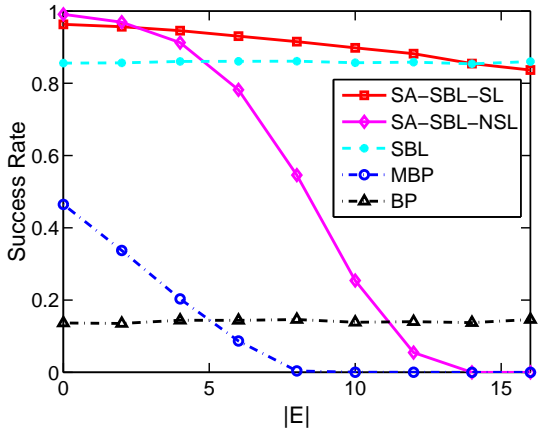}
\caption{Success rates of respective algorithms vs. the size of
the error subset.} \label{fig2}
\end{figure}

\begin{figure}[!t]
\centering
\includegraphics[width=8cm]{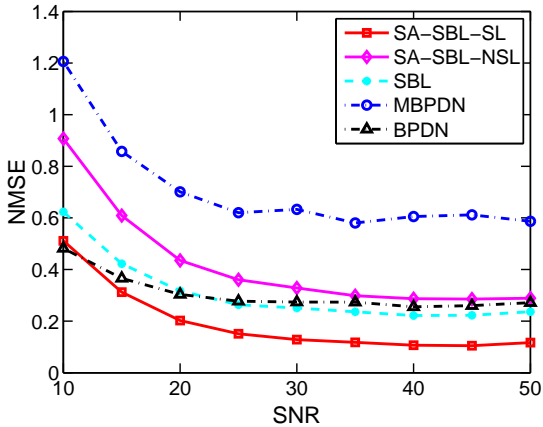}
\caption{Normalized mean-squared errors of respective algorithms
vs. the signal-to-noise ratio.} \label{fig3}
\end{figure}

\subsection{Synthetic Data}
Suppose a $K$-sparse signal is randomly generated with the support
set of the sparse signal randomly chosen according to a uniform
distribution. The signals on the support set are independent and
identically distributed (i.i.d.) Gaussian random variables with
zero mean and unit variance. The measurement matrix
$\boldsymbol{A}\in\mathbb{R}^{m\times n}$ is randomly generated
with each entry independently drawn from Gaussian distribution
with zero mean and unit variance. The prior support information
$P$ consists of two subsets: $P=S\cup E$, where $S\subset T$
denotes the subset containing the correct information, and
$E\subset T^c$ is a subset comprised of false information. In our
simulations, only the prior knowledge $P$ is available, the exact
partition of $P$ into $S$ and $E$ is unknown. We compare our
proposed algorithms with the conventional sparse Bayesian learning
(SBL), the basis pursuit (BP) method, and the modified basis
pursuit (MBP) method \cite{VaswaniLu10} which incorporates the
partial support information by assigning different
$\ell_1$-minimization weights to different coefficients.

We first consider the noiseless case. Fig. \ref{fig1} plots the
success rates of respective algorithms vs. the ratio $m/n$, where
we set $K=16$, $n=50$, $|S|=12$ and $|E|=8$, $|S|$ and $|E|$
denote the cardinality (size) of the set $S$ and $E$,
respectively. The success rate is computed as the ratio of the
number of successful trials to the total number of independent
runs. A trial is considered successful if the normalized recovery
error, i.e.
$\|\boldsymbol{x}-\boldsymbol{\hat{x}}\|_2^2/\|\boldsymbol{x}\|_2^2$,
is no greater than $10^{-6}$, where $\boldsymbol{\hat{x}}$ denotes
the estimate of the true signal $\boldsymbol{x}$. Results are
averaged over 1000 independent runs, with the measurement matrix
and the sparse signal randomly generated for each run. It can be
seen that our proposed SA-SBL-SL method presents a substantial
performance advantage over the SA-SBL-NSL and the SBL methods. The
performance gain is primarily due to the fact that the SA-SBL-SL
method is able to learn the true support from the partly erroneous
knowledge and thus make more effective use of the prior support
information. We also observe that when a considerable number of
errors are present in the prior knowledge, the methods SA-SBL-NSL
and MBP present no advantage over their respective counterparts
SBL and BP. To examine the behavior of the SA-SBL-SL method more
thoroughly, we fix the number of elements in the set $S$ and
increase the number of elements in the error set $E$. Fig.
\ref{fig2} depicts the success rates vs. the number of elements in
the error set $E$, where we set $m=25$, $K=16$, $|S|=12$ and $|E|$
varies from $1$ to $15$. As can be seen from Fig. \ref{fig2}, when
a fairly accurate knowledge is available, i.e. the number of
errors is negligible or small, the SA-SBL-NSL achieves the best
performance. This is an expected result since little learning is
required at this point. Nevertheless, as the number of elements,
$|E|$, increases, the SA-SBL-NSL suffers from substantial
performance degradation. As compared with the SA-SBL-NSL, the
SA-SBL-SL method provides stable recovery performance through
learning the values of $\{b_i\}$, and outperforms all other
algorithms when prior knowledge contains a considerable number of
errors. We, however, notice that the proposed SA-SBL-SL method is
surpassed by the conventional SBL method when inaccurate
information becomes dominant (e.g. $|E|=15$), in which case even
learning brings limited benefits and simply ignoring the
error-corrupted prior knowledge seems the best strategy.

%lacking the ability to learn the true support

%achieves a substantial performance improvement (as compared with
%the SA-SBL-NSL and SBL) through learning the true support from the
%partly erroneous knowledge.

We now consider the noisy case where the measurements are
contaminated by additive noise. The observation noise is assumed
multivariate Gaussian with zero mean and covariance matrix
$\sigma^2\boldsymbol{I}$. The normalized mean-squared errors
(NMSEs) of respective algorithms as a function of signal-to-noise
ratio (SNR) are plotted in Fig. \ref{fig3}, where we set $m=25$,
$n=50$, $K=16$, $|S|=12$, and $|E|=6$. The NMSE is calculated by
averaging normalized squared errors over $10^3$ independent runs.
The SNR is defined as $\text{SNR(dB)}\triangleq
20\log_{10}(\|\boldsymbol{Ax}\|_2/\|\boldsymbol{w}\|_2)$. The
MBP-DN is a noisy version of the MBP method \cite{LuVaswani12}. We
observe that the conventional SBL and BP-DN methods outperform
their respective counterparts: SA-SBL-NSL and MBP-DN. This, again,
demonstrates that SA-SBL-NSL and MBP-DN methods are sensitive to
prior knowledge inaccuracies. On the other hand, the proposed
SA-SBL-SL method which takes advantage of the support learning
presents superiority over both the conventional SBL as well as the
SA-SBL-NSL method.

\begin{figure}[!t]
\centering
\includegraphics[width=8cm]{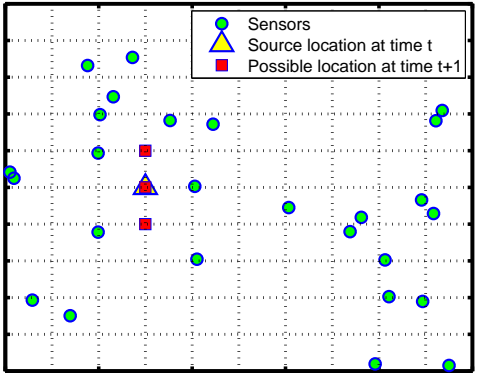}
\caption{A schematic diagram for source localization.}
\label{fig6}
\end{figure}

\begin{figure}[!t]
\centering
\includegraphics[width=8cm]{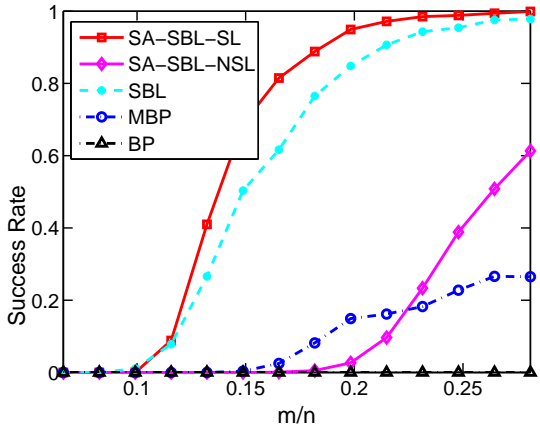}
\caption{Success rates of respective algorithms vs. the ratio
$m/n$ for the noiseless case.} \label{fig4}
\end{figure}

\begin{figure}[!t]
\centering
\includegraphics[width=8cm]{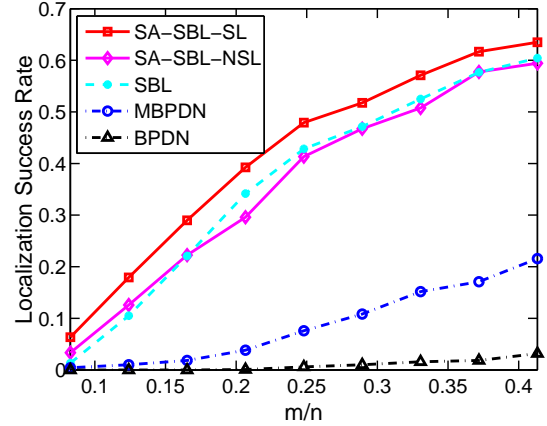}
\caption{Localization success rates of respective algorithms vs.
the ratio $m/n$, $\text{SNR}=20\text{dB}$.} \label{fig5}
\end{figure}

%estimated support in previous time step could suggest prior

\subsection{Source Localization}
We consider the problem of intensity-based source localization in
sensor networks. The sensing field is partitioned into
two-dimensional $n=121$-point virtual grid (Fig. \ref{fig6}) which
is used to represent possible locations of the $K=4$ sources. We
have $m$ randomly distributed sensors. The measurement collected
at sensors can be written as
\begin{align}
\boldsymbol{y}(t)=\boldsymbol{A}\boldsymbol{x}(t)+\boldsymbol{w}(t)
\nonumber
\end{align}
where $\boldsymbol{x}(t)\triangleq
[x_1(t)\phantom{0}\ldots\phantom{0}x_n(t)]^T$ is a sparse vector
whose entry $x_i(t)$ denotes the intensity associated with the
$i$th grid point,
$\boldsymbol{A}\triangleq[\boldsymbol{a}_1\phantom{0}\ldots\phantom{0}\boldsymbol{a}_m]^T$,
$\boldsymbol{a}_i^T\triangleq [d_{1i}^{-\alpha}\phantom{0}
d_{2i}^{-\alpha}\phantom{0} \cdots\phantom{0} d_{ni}^{-\alpha}]$,
$d_{ij}$ denotes the distance between the grid point $i$ and the
sensor $j$, and $\alpha=2$ is the energy-decay factor.

%represents the time index.
%\begin{align}
%y_j=\sum_{i=1}^{n} y_{ij}+w_j=\boldsymbol{a}_j^T\boldsymbol{x}+w_j
%\nonumber
%\end{align}
%where $w_j$ denotes the measurement noise, and
%\begin{align}
%\boldsymbol{a}_j^T\triangleq\left[d_{1j}^{-\alpha}\phantom{0}
%d_{2j}^{-\alpha}\phantom{0} \cdots\phantom{0}
%d_{nj}^{-\alpha}\right]\nonumber
%\end{align}
%The above observation model can be expressed compactly in a matrix
%form as follows

%and the following energy decay model is adopted:
%\begin{align}
%y_{ij}=\frac{x_i}{d_{ij}^{\alpha}}  \nonumber
%\end{align}
%where $y_{ij}$ denotes the signal energy measured at sensor $j$,
%$d_{ij}$ denotes the distance between the grid point $i$ and the
%sensor $j$, $x_i$ is the intensity of the $i$th source, and
%$\alpha$ is the energy-decay factor whose value is set to $2$.

%reveal some information which is

In practice, sources may keep moving but the current locations'
estimate could still be useful for the future localization. For
example, we can expect that some sources may move to grid points
close to their previous locations given that the interval between
two time instants is sufficiently small. In our simulations, we
assume that $K_1=3$ sources move slowly such that their next
locations are partially predictable based on their current
locations. Specifically, for these $K_1$ sources, each source
either stays at its current position or moves to two of its
immediate neighboring grid points\footnote{A grid point may have
more than two neighboring points but we assume the source can only
move to the specified two neighboring points, which is a
reasonable assumption in practice since objects (e.g., vehicles)
usually move along a pre-defined path such as a road or a
highway.} at the next time point (see Fig. \ref{fig6}). For
simplicity, suppose $\hat{S}=\{\hat{s}_1,\ldots,\hat{s}_{K_1}\}$
is the set of locations associated with the $K_1$ sources at time
instant $t$, then the set of possible locations of these $K_1$
sources at time instant $t+1$ is given by
$P=\{\hat{s}_1-1,\hat{s}_1,\hat{s}_1+1,\ldots,\hat{s}_{K_1}-1,\hat{s}_{K_1},\hat{s}_{K_1}+1\}$.
The set $P$ can serve as prior support knowledge for source
localization at time instant $t+1$. Nevertheless, the knowledge
$P$ is partly erroneous since all possible locations of these
$K_1$ sources in the next move are included. For the rest $K-K_1$
sources, their locations are randomly chosen from the rest $n-|P|$
grid points. To test the effectiveness of respective algorithms in
utilizing the prior knowledge, we assume the locations of these
$K_1$ sources at time instant $t$ are perfectly known to us and
examine the recovery performance at time instant $t+1$. Fig.
\ref{fig4} depicts the success rates vs. the ratio $m/n$ for the
noiseless case. Results are averaged over 1000 independent runs,
with sensors and locations of sources at time instant $t$ randomly
generated for each run. From Fig. \ref{fig4}, we see that the
SA-SBL-NSL method yields performance much worse than the
conventional SBL method. This is not surprising since the prior
knowledge contains a substantial amount of erroneous information.
It is also observed that the proposed SA-SBL-SL method which has
the ability to learn the true support from erroneous information
achieves a significant performance improvement over the SA-SBL-NSL
method. We now consider a noisy case where the measurements are
corrupted by additive Gaussian noise. When noise is present, exact
recovery of sparse signals is impossible. Nevertheless, accurate
localization can still be achieved since reliable recovery of the
support of sparse signals in the presence of noise is possible. In
our simulations, locations of sources are estimated as the grid
points associated with the largest $K$ nonzero coefficients of the
estimated signal. Fig. \ref{fig5} plots the localization success
rates as a function of the ratio $m/n$, where the signal-to-noise
ratio (SNR) is set to $20\text{dB}$. The localization success rate
is calculated as the the ratio of the number of successful trials
to the total number of independent runs. A trial is considered
successful if all $K$ sources' locations are estimated correctly.
Fig. \ref{fig5}, again, demonstrates the superiority of the
proposed SA-SBL-SL method over the SA-SBL-NSL and the conventional
SBL methods.

%measured 1 meter from the source
%locations of $K_1$ sources are temporally correlated.
%Specifically,

\begin{figure}[!t]
\centering
\includegraphics[width=8cm]{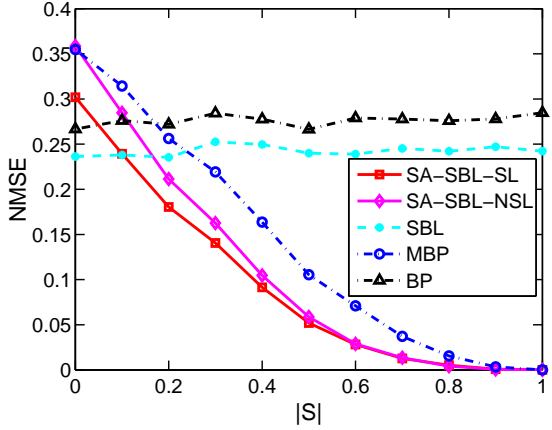}
\caption{Normalized mean squared errors of respective algorithms
vs. the cardinality of the subset $|S|$.} \label{fig7}
\end{figure}

\begin{figure}[!t]
\centering
\includegraphics[width=8cm]{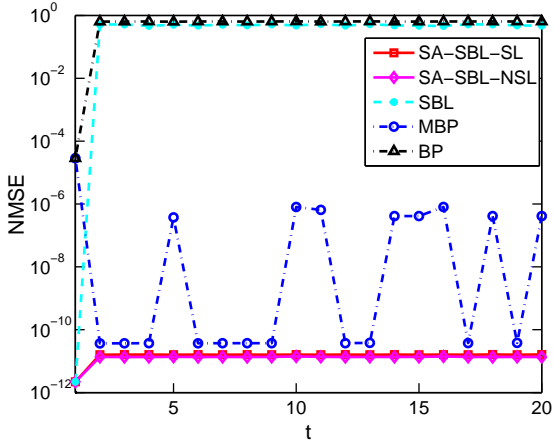}
\caption{Reconstruction of the sparsified $32\times 32$ larynx
sequence. Normalized mean squared errors vs. $t$.} \label{fig8}
\end{figure}

\subsection{MRI Data}
In this subsection, we carry out experiments on MRI images and
sequences\footnote{Available at
http://www.phon.ox.ac.uk/jcoleman/Dynamic\_MRI.html}. Images have
sparse (or approximately sparse) structures in discrete wavelet
transform (DWT) basis. By representing an image as a
one-dimensional vector, the two-dimensional DWT (a two-level
Daubechies-4 wavelet is used) of an image can be expressed as a
product of an orthonormal matrix $\boldsymbol{\Psi}$ and the image
vector $\boldsymbol{x}$. The sensing matrix $\boldsymbol{A}$ is
therefore equal to
$\boldsymbol{A}=\boldsymbol{Q}\boldsymbol{\Psi}$, where
$\boldsymbol{Q}$ denotes the measurement acquisition matrix and
its entries are i.i.d. normal random variables. We test all
algorithms on sparsified images. Similar to \cite{VaswaniLu10},
the image is sparsified by computing its 2D-DWT, retaining the
coefficients from the $99\%$-energy support while setting others
to zero and taking the inverse DWT.

We first evaluate the reconstruction performance of respective
algorithms for a sparsified image. The image is a $32\times 32$
larynx image obtained by resizing the $256\times 256$ larynx
image, i.e. $n=1024$. For this image, its support size is
$|T|=70$. In our experiments, we fix the size of the set $E$ to be
$|E|=0.4|T|$, and gradually increase the size of the set $S$. Fig.
\ref{fig7} depicts the NMSEs of respective algorithms vs. the size
of $S$, where $m=0.55n$ and $|S|$ varies from $0$ to $0.9|T|$.
Results are averaged over 500 independent runs, with the
acquisition matrix $\boldsymbol{Q}$, the sets $E$ and $S$ randomly
generated for each run. From Fig. \ref{fig7}, we see that when the
prior support knowledge is dominated by the error set $E$, the SBL
and BP methods outperform their respective ``support-aided''
counterparts SA-SBL-NSL and MBP. Nevertheless, the SA-SBL-NSL and
the MBP methods surpass and eventually achieve a significant
performance improvement over the SBL and BP methods as the prior
knowledge becomes more and more accurate. It can also be observed
the SA-SBL-SL method presents uniform superiority over SA-SBL-NSL
and MBP for different values of $|S|$, and the performance gap is
wider in the small $|S|$'s region since learning in this region
apparently brings more significant benefits as compared with
learning in the large $|S|$'s region.

%and achieve a significant performance improvement over the SBL and
%BP methods when a fairly accurate prior knowledge is available

We also conduct a comparison for the sparsified $32\times 32$
larynx sequence in Fig. \ref{fig8}. Since the support of the
sequence undergoes a small variation, the prior support knowledge
can be obtained as the estimate of the previous time instant. At
the very beginning, i.e. $t=0$, the prior knowledge set $P$ is
empty. Conventional SBL and BP methods are then used to obtain an
initial support estimate for their respective support-aided
methods. We set $m_0=0.6n$ in order to ensure a fairly accurate
initial estimate is obtained. For $t>0$, only $m_t=0.3n$
measurements are collected for the signal reconstruction. From
Fig. \ref{fig8}, we see that the all support-aided methods
including SA-SBL-SL, SA-SBL-NSL and MBP are able to achieve exact
reconstruction with only as few as $m=0.3n$ measurements, whereas
the conventional SBL and BP fail to recover the signal with these
few measurements. Also, since the prior support knowledge is
fairly accurate, support learning does not bring any additional
benefits, and thus both the SA-SBL-SL and SA-SBL-NSL methods
attain similar recovery performance.

%we resort to  as the prior set $P$ is empty
%to obtain an initial estimate of the image, where we set $m=0.5n$.
%The initial estimate then serves as prior support knowledge in the

%beneficial than learning for the case where the prior knowledge is
%fairly accurate.

\section{Conclusions} \label{sec:conclusion}
We studied the problem of sparse signal recovery given that part
of the signal's support is known \emph{a priori}. The prior
knowledge, however, may not be accurate and could contain
erroneous information. This knowledge inaccuracy may result
considerable performance degradation or even recovery failure. To
address this issue, we first introduced a modified two-layer
Gaussian-inverse Gamma hierarchical prior model. The modified
two-layer model employs an individual parameter $b_i$ for each
sparsity-controlling hyperparameter $\alpha_i$, and therefore has
the ability to place non-sparsity-encouraging priors to those
coefficients that are believed in the support set. Based on this
two-layer mode, we then proposed an improved three-layer
hierarchical prior model, with a prior placed on the parameters
$\{b_i\}$ in the third layer. Such a model enables to
automatically learn the true support from partly erroneous
information through learning the values of $\{b_i\}$. Bayesian
algorithms are developed by resorting to the mean field
variational Bayes. Simulation results show that substantial
performance improvement can be achieved through support learning
since it allows us to make more effective use of the partly
erroneous information.

\end{document}